\documentclass[a4paper,11pt]{article}

\usepackage[T1]{fontenc}
\usepackage{times}
\usepackage{graphicx}
\usepackage{amsfonts}
\usepackage{amssymb}
\usepackage{amsmath} 
\usepackage{amsthm}
\usepackage{natbib}
\usepackage{graphicx}
\usepackage{pdfpages}
\usepackage{tabularx}
\usepackage{hyperref}

\usepackage[margin=1in]{geometry}

\usepackage{pifont}
\makeatletter
\def\blfootnote{\xdef\@thefnmark{}\@footnotetext}
\makeatother

\newcommand {\RR}{\mathbb{R}}
\newcommand {\OMIT}[1]{}

\begin{document}

\title{Large-scale Machine Learning for Metagenomics Sequence Classification}
\author{K\'evin Vervier\,$^{1,2,3,4}$, Pierre Mah\'e\,$^{1,*}$, Maud Tournoud\,$^{1}$,\\Jean-Baptiste\ Veyrieras\,$^{1}$ and Jean-Philippe Vert\,$^{2,3,4}$}

\maketitle
\blfootnote{\noindent $^{1}$Bioinformatics Research Departement, bioM\'erieux, 69280 Marcy-l'\'Etoile, France\\
$^{2}$MINES ParisTech, PSL Research University, CBIO-Centre for Computational Biology, 77300 Fontainebleau, France\\
$^{3}$ Institut Curie, 75248 Paris Cedex, France, \\$^{4}$ INSERM U900, 75248 Paris Cedex, France\\
$^*$ contact: \href{mailto:pierre.mahe@biomerieux.com}{pierre.mahe@biomerieux.com}}
\begin{abstract}

Metagenomics characterizes the taxonomic diversity of microbial communities by sequencing DNA directly from an environmental sample. One of the main challenges in metagenomics data analysis is the binning step, where each sequenced read is assigned to a taxonomic clade. Due to the large volume of metagenomics datasets, binning methods need fast and accurate algorithms that can operate with reasonable computing requirements. While standard alignment-based methods provide state-of-the-art performance, compositional approaches that assign a taxonomic class to a DNA read based on the $k$-mers it contains have the potential to provide faster solutions.

In this work, we investigate the potential of modern, large-scale machine learning implementations for taxonomic affectation of next-generation sequencing reads based on their $k$-mers profile. We show that machine learning-based compositional approaches benefit from increasing the number of fragments sampled from reference genome to tune their parameters, up to a coverage of about 10, and from increasing the $k$-mer size to about 12. Tuning these models involves training a machine learning model on about $10^8$ samples in $10^7$ dimensions, which is out of reach of standard softwares but can be done efficiently with modern implementations for large-scale machine learning. The resulting models are competitive in terms of accuracy with well-established alignment tools for problems involving a small to moderate number of candidate species, and for reasonable amounts of sequencing errors.
We show, however, that compositional approaches are still limited in their ability to deal with problems involving a greater number of species, and more sensitive to sequencing errors.
We finally confirm that compositional approach achieve faster prediction times, with a gain of 3 to 15 times with respect to the BWA-MEM short read mapper, depending on the number of candidate species and the level of sequencing noise.



\end{abstract}

\section{Introduction}


Recent progress in next-generation sequencing (NGS) technologies allow to access large amounts of genomic data within a few hours at a reasonable cost~\citep{soon2013high}.  In metagenomics, NGS is used to analyse the genomic content of microbial communities by sequencing all DNA present in an environmental sample~\citep{riesenfeld2004metagenomics}. It gives access to all organisms present in the sample even if they do not grow on culture media \citep{hugenholtz2002exploring}, and allows us to characterize with an unprecedented level of resolution the diversity of the microbial realm \citep{peterson2009nih}.

The raw output of a metagenomics experiment is a large set of short DNA sequences (reads) obtained by high-throughput sequencing of the DNA present in the sample. There exist two main approaches to analyze these data, corresponding to slightly different goals. On the one hand, {\it taxonomic  profiling} aims to estimate the relative abundance of the members of the microbial community, without necessarily affecting each read to a taxonomic class. Recent works like WGSQuikr \citep{koslicki2014wgsquikr} or GASIC \citep{lindner2012metagenomic} proved to be very efficient for this purpose. {\it Taxonomic binning} methods, on the other hand, explicitly affect each read to a taxonomic clade. This process can be unsupervised, relying on clustering methods to affect reads to operational taxonomic units (OTU), or supervised, in which case reads are individually affected to nodes of the taxonomy~\citep{mande2012classification}. While binning is arguably more challenging that profiling, it is a necessary step for downstream applications which require draft-genome reconstruction. This may notably be the case in a diagnostics context, where further analyses could aim to detect pathogen micro-organisms~\citep{miller2013} or antibiotic resistance mechanisms~\citep{schmieder2012insights}. 

In this paper we focus on the problem of supervised taxonomic binning, where we wish to assign each read in a metagenomics sample to a node of a pre-defined taxonomy. Two main computational strategies have been proposed for that purpose: (i) alignment-based approaches, where the read is searched against a reference sequence database with sequence alignment tools like BLAST \citep{huson2007megan} or short read mapping tools \citep[e.g., BWA,][]{li2009fast}, and (ii) compositional approaches, where a machine learning model such as a naive Bayes (NB) classifier \citep{wang2007naive,parks2011} or a support vector machine \citep[SVM, ][]{mchardy2007accurate,patil2012phylopythias} is trained to label the read based on the set of $k$-mers it contains. Since the taxonomic classification of a sequence by compositional approaches is only based on the set of $k$-mers it contains, they can offer significant gain in terms of classification time over similarity-based approaches. Training a machine learning model for taxonomic binning can however be computationally challenging. Indeed, compositional approaches must be trained on a set of sequences with known taxonomic labels, typically obtained by sampling error-free fragments from reference genomes. In the case of NB classifiers, explicit sampling of fragments from reference genomes is not needed to train the model: instead, a global profile of $k$-mer abundance from each reference genome is sufficient to estimate the parameters of the NB model, leading to simple and fast implementations \citep{wang2007naive,rosen2011nbc,parks2011}. On the other hand, in the case of SVM and related discriminative methods, an explicit sampling of fragments from reference genomes to train the model based on the $k$-mer content of each fragment is needed, which can be a limitation for standard SVM implementations. For example, \cite{patil2012phylopythias} sampled approximately 10,000 fragments from 1768 genomes to train a structured SVM (based on a $k$-mer representation with $k=4,5,6$), and reported an accuracy competitive with similarity-based approaches. Increasing the number of fragments sampled to train a SVM may improve its accuracy, and allow us to investigate larger values of $k$. However it also raises computational challenges, as it involves machine learning problems where a model must be trained from potentially millions or billions of training examples, each represented by a vector in $10^7$ dimensions for, e.g., $k=12$.

In this work, we investigate the potential of compositional approaches  for taxonomic label assignment using modern, large-scale machine learning algorithms. 

\OMIT{
We first further detail our motivation and introduce the machine learning  algorithm considered in Section \ref{sec:classif}.
We then carry out a proof of concept on a toy dataset in Section \ref{sec:poc} demonstrating the feasibility of the approach, and the necessity to consider such large scale machine learning methods.
Section \ref{sec:results} describes an extensive evaluation on a realistic dataset involving several hundreds of bacterial species, investigating in particular the robustness of compositional approaches to sequencing errors.
We finally discuss these results and perspectives for future work in Section \ref{sec:discussion}.
}

\section{Methods}

\subsection{Linear models for read classification}\label{sec:classif}

In most of compositional metagenomics applications, a sequence is represented by its $k$-mer profile, namely, a vector counting the number of occurrences of any possible word of $k$ letters in the sequence. Only the $A, T, C ,G$ nucleotides are usually considered to define $k$-mer profiles, that are therefore $4^k$-dimensional vectors. Although the size of the $k$-mer profile of a sequence of length $l$ increases exponentially with $k$, it contains at most $l-k+1$ non-zero elements since a sequence of length $k$ contains $l-k+1$ different $k$-mers.

Given a sequence represented by its $k$-mer profile $x\in\RR^{4^k}$, we consider linear models to assign it to one of $K$ chosen taxonomic classes. A linear model is a set of weight vectors $w_1,\ldots,w_K \in \RR^{4^k}$ that assign $x$ to the class
$$
\arg\max_{j=1,\ldots,K} w_j^\top x\,,
$$
where $w^\top x$ is the standard inner product between vectors. To train the linear model, we start from a training set of sequences $x_1,\ldots,x_n \in \RR^{4^k}$ with known taxonomic labels $c_1,\ldots,c_n \in \left\{1,\ldots,K\right\}$. A NB classifier, for example, is a linear model where the weights are estimated from the $k$-mer count distributions on each class. Another class of linear models popular in machine learning, which include SVM, are the discriminative approaches that learn the weights by solving an optimization problem which aims to separate the training data of each class from each other. More precisely, to optimize the weight $w_j$ of the $j$-th class, one typically assigns a binary label $y_i$ to each training example ($y_i=1$ if $c_i=j$, or $y_i=-1$ otherwise) and solves an optimization problem of the form
\begin{equation}\label{eq:learning}
\min_{w} \frac{1}{n} \sum_{i=1}^n \ell(y_i, w^\top x_i) + \lambda \|w\|^2 \,,
\end{equation}
where $\ell(y,t)$ is a loss function quantifying how "good" the prediction $t$ is if the true label is $y$, and $\lambda\geq 0$ is a regularization parameter to tune, helpful to prevent overfitting in high dimension. A SVM solves (\ref{eq:learning}) with the hinge loss $\ell(y,t) = max(0,1-yt)$, but other losses such as the logistic loss $\ell(y,t) = \log(1+\exp(-yt))$ or the squared loss $\ell(y,t)=(y-t)^2$ are also possible and often lead to models with similar accuracies. These models have met significant success in numerous real-world learning tasks, including compositional metagenomics \citep{patil2012phylopythias}. In this work, we use the squared loss function and choose $\lambda = 0$, leading to no regularization.

\subsection{Large-scale learning of linear models}
Although learning linear models by solving (\ref{eq:learning}) is now a mature technology implemented in numerous softwares, metagenomics applications raise computational challenges for most standard implementations, due to the large values that $n$ (number of reads in the training set), $p=4^k$ (dimension of the models) and $K$ (number of taxonomic classes) can take.

The training set is typically obtained by sampling fragments from reference genomes with known taxonomic class. For example, \citet{patil2012phylopythias} sampled approximately $n=10,000$ fragments from $1,768$ genomes to train SVM models based on $k$-mer profiles of size $k=4,5,6$. However, the number of distinct fragments that may be drawn from a genome sequence is approximately equal to its length (by sampling a fragment starting at each position in the genome), hence can reach several millions for each microbial genome, leading to potentially billions of training sequences when thousands of reference genomes are used. While considering every possible fragment from every possible genome may not be the best choice because of the possible redundancy between the reads, it may still be useful to consider a significant number of fragments to properly account for the intra and inter species genomic variability. Similarly, exploring models with $k$ larger than $6$, say $10$ or $15$, may be interesting but requires (i) the capacity to manipulate the corresponding $4^k$-dimensional vectors ($4^{15} \sim 10^9$), and (ii) large training sets since many examples are needed to learn a model in high dimension. Finally, real-life applications involving actual environmental samples may contain several hundreds microbial species, casting the problem into a relatively massive multiclass scenario out of reach of most standard implementations of SVM.

To solve (\ref{eq:learning}) efficiently when $n$, $k$ and $K$ take large values, we use a dedicated implementation of stochastic gradient descent~\citep[SGD][]{bottou1998online} available in the Vowpal Wabbit software~\citep[VW][]{langford2007vowpal,JMLR:v15:agarwal14a}. In short, SGD exploits the fact that the objective function in (\ref{eq:learning}) is an average of $n$ terms, one for each training example, to approximate the gradient at each step using a single, randomly chosen term. Although SGD requires more steps to converge to the solution than standard gradient descent, each step is $n$ times faster and the method is overall faster and more scalable. In addition, although the dimension $p=4^k$ of the data is large, VW exploits the fact that each training example is sparse, leading to efficient memory storage and fast updates at each SGD step. We refer the interested reader to \citet{bottou2010large} for more discussion about the relevance of SGD in large-scale learning. In practice, VW can train a model with virtually no limit on $n$ as long as the data can be stored on a disk (they are not loaded in memory). As for $k$, VW can handle up to $2^{32}$ distinct features, and the count of each $k$-mer is randomly mapped to one feature by a hash table. This means that we have virtually no limit on $k$, except that when $k$ approaches or exceeds the limit (such that $4^k = 2^{32}$, i.e., $k=16$), collisions will appear in the hash table and different $k$-mers will be counted together, which may impact the performance of the model.

\section{Data}

We simulate metagenomics samples by generating reads from three different reference databases, which we refer to below as the \emph{mini}, the \emph{small} and the \emph{large} databases.

The \emph{mini} reference database contains 356 complete genome sequences covering 51 bacterial species, listed in Table \ref{Tab:poc}. We use this database to train and extensively vary the parameters of the different models. To measure the performance of the different models, we generate new fragments from 52 genomes not present in the reference database, but originating from one of the 51 species\footnote{Two genomes are indeed available for the {\it Francisella tularensis} species, one of which originating from the {\it novicida} subspecies.}.
\begin{table}[!t]
\begin{center}
\begin{tabular}{l l}
\hline
\textit{Acetobacter pasteurianus} & \textit{Methylobacterium extorquens}\\ 
\textit{Acinetobacter baumannii} &  \textit{Mycobacterium bovis}\\  
\textit{Bacillus amyloliquefaciens} & \textit{Mycobacterium tuberculosis}  \\ 
\textit{Bacillus anthracis} &  \textit{Mycoplasma fermentans}\\
\textit{Bacillus subtilis }& \textit{Mycoplasma genitalium} \\ 
\textit{Bacillus thuringiensis} &  \textit{Mycoplasma mycoides} \\ 
\textit{Bifidobacterium bifidum} & \textit{Mycoplasma pneumoniae} \\ 
\textit{Bifidobacterium longum} & \textit{Neisseria gonorrhoeae}  \\ 
\textit{Borrelia burgdorferi} &\textit{Propionibacterium acnes}  \\  
\textit{Brucella abortus} & \textit{Pseudomonas aeruginosa}\\ 
\textit{Brucella melitensis} & \textit{Pseudomonas stutzeri}  \\ 
\textit{Buchnera aphidicola} &\textit{Ralstonia solanacearum} \\ 
\textit{Burkholderia mallei} &  \textit{Rickettsia rickettsii} \\ 
\textit{Burkholderia pseudomallei} & \textit{Shigella flexneri} \\ 
\textit{Campylobacter jejuni} & \textit{Staphylococcus aureus} \\ 
\textit{Corynebacterium pseudotuberculosis} & \textit{Streptococcus agalactiae}\\ 
\textit{Corynebacterium ulcerans} &  \textit{Streptococcus equi} \\ 
\textit{Coxiella burnetii} & \textit{Streptococcus mutans} \\  
\textit{Desulfovibrio vulgaris} & \textit{Streptococcus pneumoniae} \\ 
\textit{Enterobacter cloacae} & \textit{Streptococcus thermophilus} \\ 
\textit{Escherichia coli}  &\textit{Thermus thermophilus}  \\ 
\textit{Francisella tularensis} & \textit{Treponema pallidum} \\ 
\textit{Helicobacter pylori} & \textit{Yersinia enterocolitica} \\ 
\textit{Legionella pneumophila} &\textit{Yersinia pestis} \\ 
\textit{Leptospira interrogans} &  \textit{Yersinia pseudotuberculosis}  \\ 
\textit{Listeria monocytogenes}  & \\
\hline
\end{tabular}
\end{center}
\caption{List of the 51 microbial species in the \emph{mini} reference database.\label{Tab:poc}}
\end{table}

The \emph{small} and \emph{large} databases are meant to represent more realistic situations, involving a larger number of candidate bacterial species and a larger number of reference genomes.
To define the reference and validation databases that will respectively be used to build and evaluate the predictive models, we first downloaded the 5201 complete bacterial and archeal genomes available in the NCBI RefSeq database as of July 2014 \citep{pruitt2012ncbi}, by means of a functionality embedded in the Fragment Classification Package (FCP) \citep{parks2011}.
We then filtered these sequences according to a criterion proposed in \citet{parks2011}: we only kept genomes that belong to genera represented by at least 3 species. 
We also removed genomes represented by less than $10^6$ nucleotides in order to filter draft genome sequences, plasmids, phages, contigs and other short sequences. 
The 2961 remaining sequences originate from 774 species, among which 193 are represented by at least 2 strains. 
We split the sequences of these 193 species into two parts. 
We randomly pick one strain within each of these 193 species to define a validation database, that will be used to estimate classification  performance, through the sampling of genomic fragments or the simulation of sequencing reads.
The remaining sequences of these 193 species define a first reference database, referred to as {\it small} below. 
In addition we define a larger reference database by adding to the {\it small} database described above the genomes originating from the $774-193=581$ species represented by a single genome. The larger database, referred to as {\it large} below, therefore involves the 774 species available after filtering the NCBI database, and not solely the 193 represented in the validation database.

\section{Results}
\subsection{Proof of concept on the \emph{mini} database}\label{sec:poc}
In this section we present a study on the \emph{mini} dataset, aiming to evaluate the impact of increasing the number of fragments used to train the model as  well as the length of the $k$-mers considered.
For that purpose, we learn several classification models based on fragments of length $L=200$ or $L=400$ sampled from the 356 reference genomes in the \emph{mini} reference database, represented by $k$-mers of size in $\{4,6,8,10,12\}$.
The  number of fragments used to learn the models is gradually increased by drawing several "batches" of fragments in order to cover, on average, each nucleotide of the reference genomes a pre-defined number of times $c$. We vary the coverage $c$ between  $0.1$ to its maximal value, equal to the length of the fragments considered.  This leads to learning models from around $n=2.7\times10^5$, for $c=0.1$ and $L=400$, up to around $n=1.1\times10^9$ fragments, when $c$ reaches its maximal value. This is way beyond the configurations considered for instance in \citet{patil2012phylopythias}, where SVM  models were learned from approximately $10^4$ fragments drawn from $1,768$ genomes.

To assess the performance of these models we consider two sets of $134,319$  fragments, of respective length 200 and 400, drawn from the 52 complete genomes that are not in the reference database used to train the models. Performance is measured by first computing, for each species, the proportion of fragments that are correctly classified, and considering its median value across species. In a multiclass setting, this indicator is indeed less biased towards over-represented classes than the global rate of correct classification.

\begin{figure}[!tpb]
	\centerline{\includegraphics[page=1,width=0.5\textwidth]{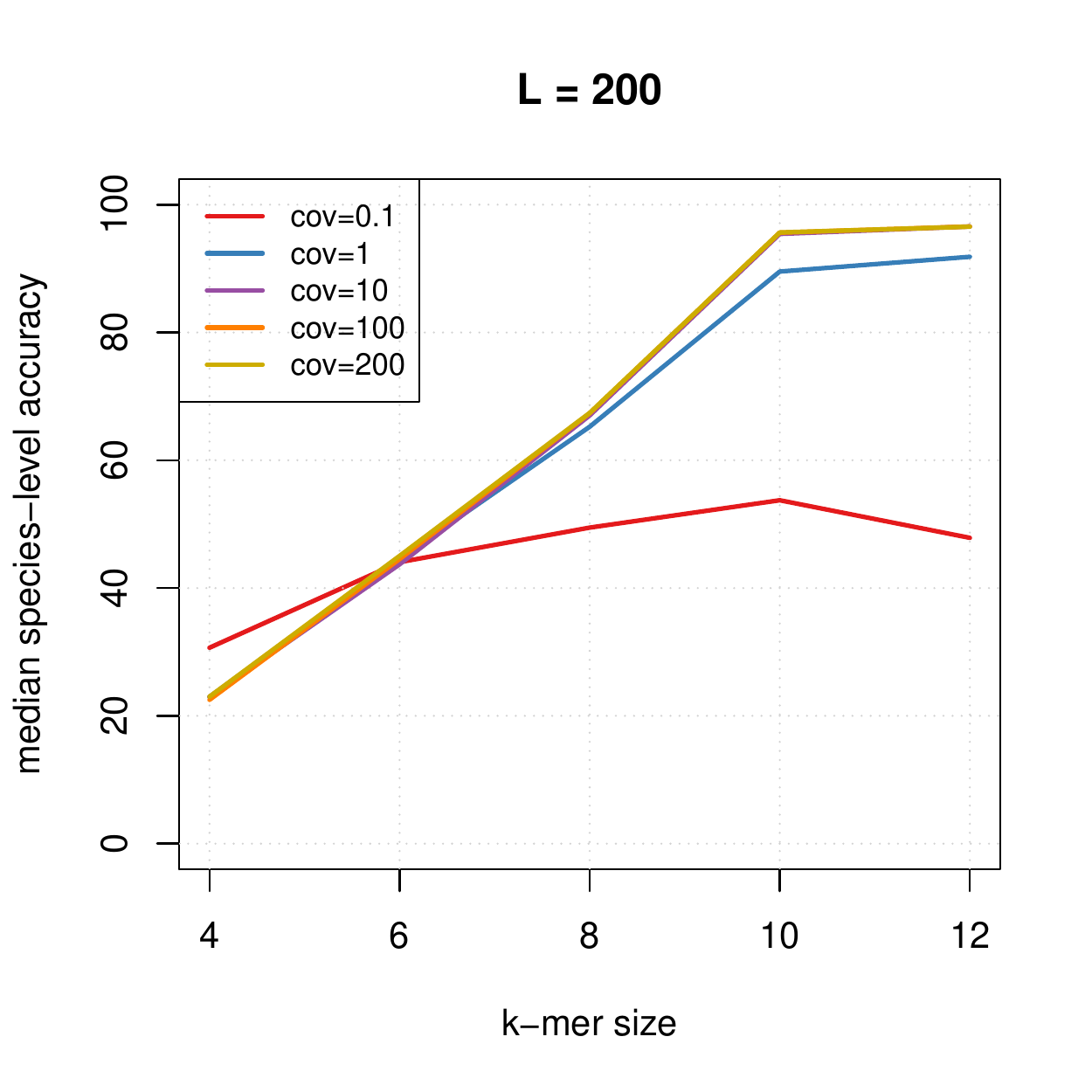}
		\includegraphics[page=1,width=0.5\textwidth]{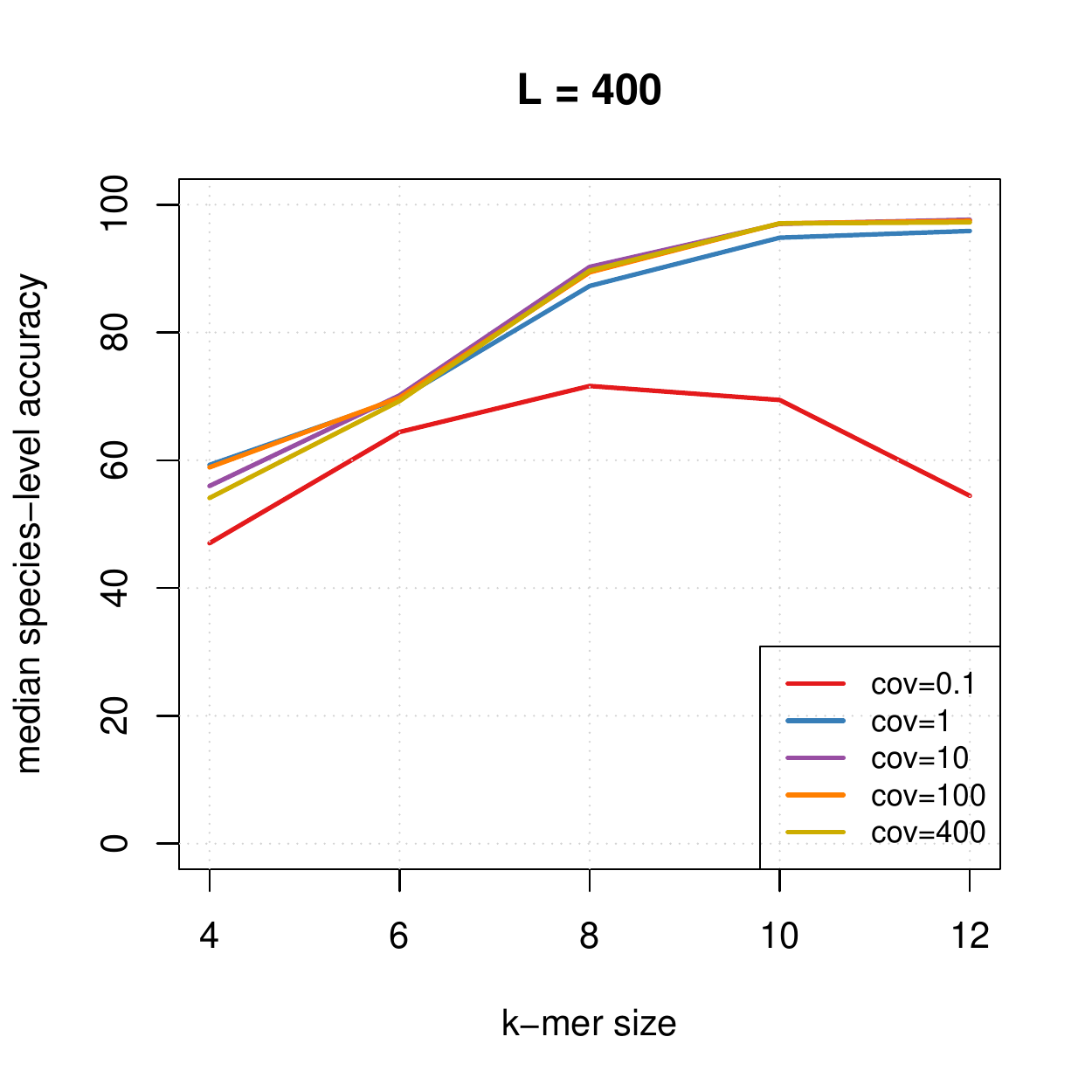}}
	{\caption[\textbf{Increasing the number of fragments and the $k$-mer size on the \textit{mini} datasets}.]{\textbf{Increasing the number of fragments and the $k$-mer size on the \textit{mini} datasets}. Left: $L=200$bp fragments. Right: $L=400$bp fragments. These figures show the median species-level accuracy obtained by linear predictors trained with Vowpal Wabbit from fragments covering each reference genome with a mean coverage $c$ from 0.1 (red) to $L$ (gold). Performances are reported as a function of $k$-mer sizes.}\label{fig:covmax}}
\end{figure}

Figure \ref{fig:covmax} shows the performance reached by models based on fragments of length 200 (left) or 400 (right), for different values of $k$ (horizontal axis) and different coverages (different colors).
We first note that for $c=0.1$, that is, for a limited number of fragments, the classification performance starts by increasing with the size of the $k$-mers (up to $k=8$ and $k=10$ for fragments of length 200 and 400, respectively), and subsequently decreases. This suggests that the number of fragments considered in this setting is not sufficient to efficiently learn when the dimensionality of the feature space becomes too large. Note that twice as many fragments of length 200 as fragments of size 400 are drawn for a given coverage value, which may explain why performance still increases beyond $k=8$ with smaller fragments.
Increasing the number of fragments confirms this hypothesis : performance systematically increases or remains steady with $k$ for $c \geq1$, and for $k\geq 8$, the performance is significantly higher than that obtained at $c=0.1$, for both length of fragments.
Increasing the coverage from $c=1$ to $c=10$  has a positive impact in both cases, although marginally for fragments of length 400. Further increasing the number of fragments does not bring any improvement.

Altogether, the optimal configuration on this {\it mini} dataset involves $k$-mers of size 12 and drawing fragments at a coverage $c\geq10$ for the two lengths of fragments considered. Further increasing the size of the $k$-mers did not bring  improvements, and actually proved to be challenging. 
Indeed, as mentioned above, VW proceeds by hashing the input features into a vector offering at most $2^{32}$ entries.
This hashing operation can induce collisions between features, which can be detrimental to the model if the number of features becomes too high with respect to the size of the hash table. This issue is even more stressed in a multiclass setting, where the number of hash table entries available per model is divided by the number of classes considered. On this dataset, 51 models have to be stored in the hash table, which reduces the number of entries available per model to $2^{32}/51 \sim 2^{32-6} = 4^{13}$. We have empirically observed that performance could not increase for $k$ greater than 12 and actually decreased for $k$-mers greater than 15. 

\OMIT{
This issue is further discussed and illustrated in Supplementary materials, Figure \ref{fig:large-k}.

\begin{figure*}[!tpb]
	\centerline{\includegraphics[page=2,width=0.85\textwidth]{results-frag-200_large-k.pdf}}
	{\caption[\textbf{Large $k$-mer sizes and collisions in hash table.}]{\textbf{Large $k$-mer sizes and collisions in hash table.} Left: Median species-accuracy. Middle: Mean species accuracy. Right: Micro-accuracy. These figures give classification accuracy as a function of the $k$-mer size and the mean coverage.
		}\label{fig:large-k}}
\end{figure*}
}

\begin{figure}[!tpb]
	\centerline{\includegraphics[page=2,width=0.5\textwidth]{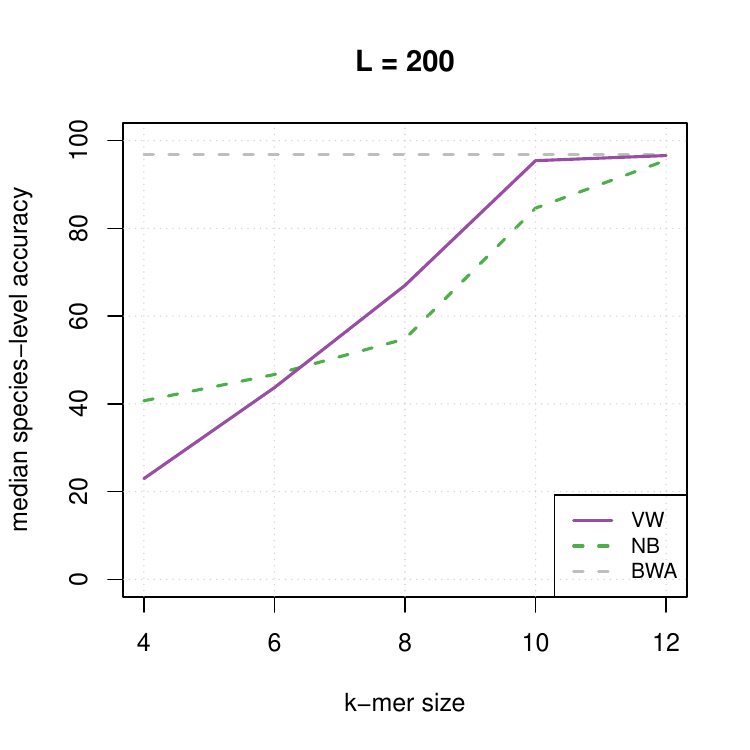}
		\includegraphics[page=2,width=0.5\textwidth]{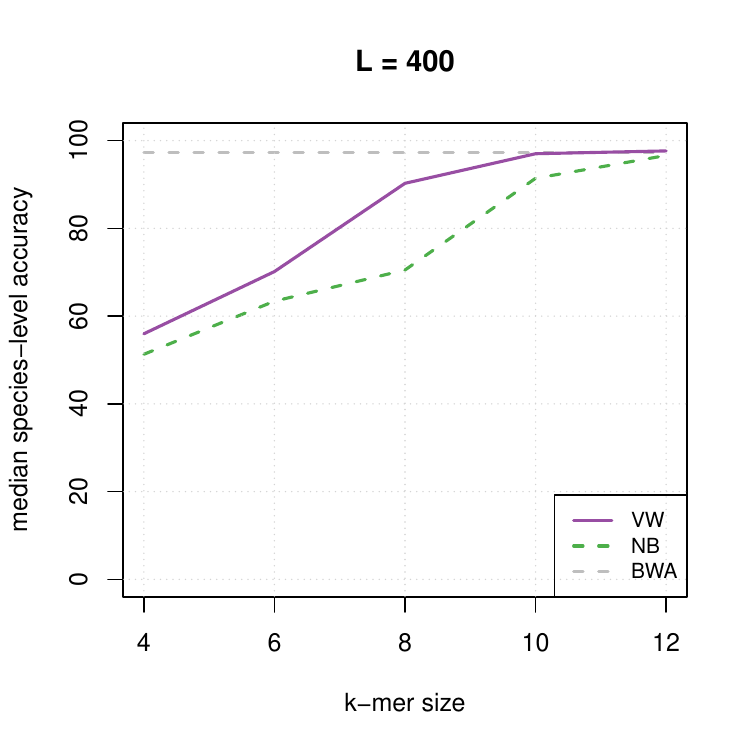}}
	{\caption[\textbf{Comparison between Vowpal Wabbit and reference methods on the {\it mini} datasets.}]{\textbf{Comparison between Vowpal Wabbit and reference methods on the {\it mini} datasets.} Left: $L= 200$bp fragments. Right: $L=400$bp fragments. These figures show the median species-level accuracy obtained by linear predictors trained with Vowpal Wabbit from fragments covering each reference genome with a mean coverage equal to $10$ (purple solid line). Performances are reported as a function of $k$-mer sizes. This approach is compared to the standard compositional Naive Bayes approach (green dotted line), and an alignment-based approach based on BWA (grey dotted line).
		}\label{fig:NB}}
\end{figure}

We now compare these results to two well-established approaches: a comparative approach based on the BWA-MEM sequence aligner \citep{li2013aligning} and a compositional approach based on the generative NB classifier \citep{rosen2011nbc}.
The NB experiments rely on the FCP implementation \citep{parks2011} and are carried out in the same setting as VW: we compute profiles of $k$-mers abundance for the 356 genomes of the reference database, and use them to affect test fragments to their most likely genome.
BWA-MEM is configured to solely return hits with maximal score (option \texttt{-T 0}). 
Unmapped fragments are counted as misclassifications, and a single hit is randomly picked in case of multiple hits, in order to obtain a species-level prediction. This latter random hit selection process is repeated 20 times and the performance indicator reported below corresponds to its median value obtained across repetitions.
Results are shown in Figure \ref{fig:NB}.
We first note that $k$-mer based approaches, either generative or discriminative, never outperform the alignment-based approach.
Comparable results are nevertheless obtained for $k \geq 10$ with VW, and $k=12$ with the NB. 
Performances obtained for shorter $k$-mers are markedly lower than that obtained by BWA-MEM.
We note finally that VW generally outperforms the NB classifier, except for small $k$-mers and short fragments ($k\leq6$ and $L=200$).

In summary, these experiments demonstrate the relevance and feasibility of large-scale machine learning for taxonomic binning: we obtain a performance comparable to that of the well-established alignment-based approach, provided a sufficient number of fragments and long enough $k$-mers are considered to learn the $k$-mers based predictive models.


\subsection{Evaluation on the \emph{small} and \emph{large} reference databases}
We now proceed to a more realistic evaluation involving a larger number of candidate microbial species and a larger number of reference genomes, using the \emph{small} and \emph{large} reference databases.
We learn classification models according to the configuration suggested by the evaluation on the \emph{mini} database: we consider $k$-mers of size 12 and a number of fragments allowing to cover each base of the reference genomes 10 times in average. We limit our analysis to fragments of length 200, which leads to learn models from around $n=1.38\times 10^8$ and $n=2.56\times 10^8$ fragments for the small and large reference databases, respectively.
Note that due to the larger number of species involved, around $2^{32-8}=4^{12}$ and $2^{32-10}=4^{11}$ entries of the VW hash table are available per model for each of these reference databases. Based on the results with the \emph{mini} database, this should be compatible with $k$-mers of size 12.  
We evaluate the performance of the models on fragments extracted from the 193 genomes of the validation database and draw a number of reads necessary to cover each base of each genome once in average, which represents around $3.5\times 10^6$ sequences. Results obtained by VW are again compared to that obtained by the two baseline approaches involved in the previous proof of concept, namely BWA-MEM and NB, and are shown in Table \ref{tab:FCP-referenceDB}.
We first note that for the \emph{small} reference database, the performance of VW and BWA-MEM are very similar (median species-level accuracy of 92.4\% and 93\%, respectively). The NB classifier, on the other hand, has a significantly lower performance, with 8\% less in median accuracy than the alignment-based approach. Considering a larger number of candidate species in the \emph{large} reference database has little impact on the alignment-based approach, where we observe a performance drop of only 1\%  (91.9\% vs 93\%). It impacts more severely compositional approaches, with both NB and VW accuracies dropping by about 5\%. This suggests that $k$-mer based approach are still limited in their ability to deal with problems involving more than a few hundreds of candidate species.
\OMIT{The generative NB approach was already outperformed by the alignment-based approach for 193 species, while the discriminative VW approach remained competitive.
Both approaches showed however significantly lower performances for 774 species.
}
\OMIT{
\begin{figure}[!tpb]
	\centerline{\includegraphics[page=1,width=0.45\textwidth]{FCP-fragments.pdf}}
	{\caption[\textbf{Evaluation on FCP dataset: comparison between reference databases.}]{\textbf{Evaluation on FCP dataset: comparison between reference databases.} Top: smallDB reference. Bottom: largeDB reference. These figures give median accuracy by taxon for VW (purple), NB (green) and  BWA (grey). Compositional approaches (VW and NB) performances are reported for a kmer length $k$ equal to 12. VW performances are reported for a mean coverage $c=10$. \todo{should we use a table instead ?}
		}\label{fig:FCP-referenceDB}}
\end{figure}
}
\begin{table}[!t]
\begin{center}
\begin{tabular}{lcc}\hline
 & small database & large database \\\hline
Vowpal Wabbit & 92.4 & 87.7\\
Naive Bayes & 85.1 & 79.8\\
BWA-MEM & 93.0 & 91.9\\\hline
\end{tabular}
\end{center}
\caption{\textbf{Performance on the \emph{small} and \emph{large} reference databases.} This table gives the median  species-level accuracy obtained by Vowpal Wabbit (VW), Naive Bayes (NB) and BWA-MEM using the small and large reference databases. Compositional approaches (VW and NB) performances are reported for $k$-mers of size 12. VW performances are reported for a mean coverage $c=10$.\label{tab:FCP-referenceDB}}
\end{table}

\subsection{Robustness to sequencing errors}
The evaluation performed in the previous sections is based on taxonomic classification of DNA fragments drawn from reference genomes without errors. In real life, sequencing errors may alter the read sequences and make the classification problem more challenging. To evaluate the robustness of the classifiers to sequencing errors, we generate new reads simulating sequencing errors using the Grinder read simulation software \citep{angly2012grinder}. We consider two types of sequencing errors models:  homopolymeric stretches, which are commonly encountered in pyrosequencing technologies (e.g., Roche 454), and general mutations (substitutions and insertions/deletions). In order to be able to compare the results of the fragment- and read-based evaluations, we systematically simulate reads of length 200 (exactly), and simulate around $3.5\times 10^6$ sequences as well. 

\subsubsection*{Homopolymeric error models.}
\begin{figure}[!tpb]
	\centerline{\includegraphics[page=1,width=\textwidth ,height=8cm]{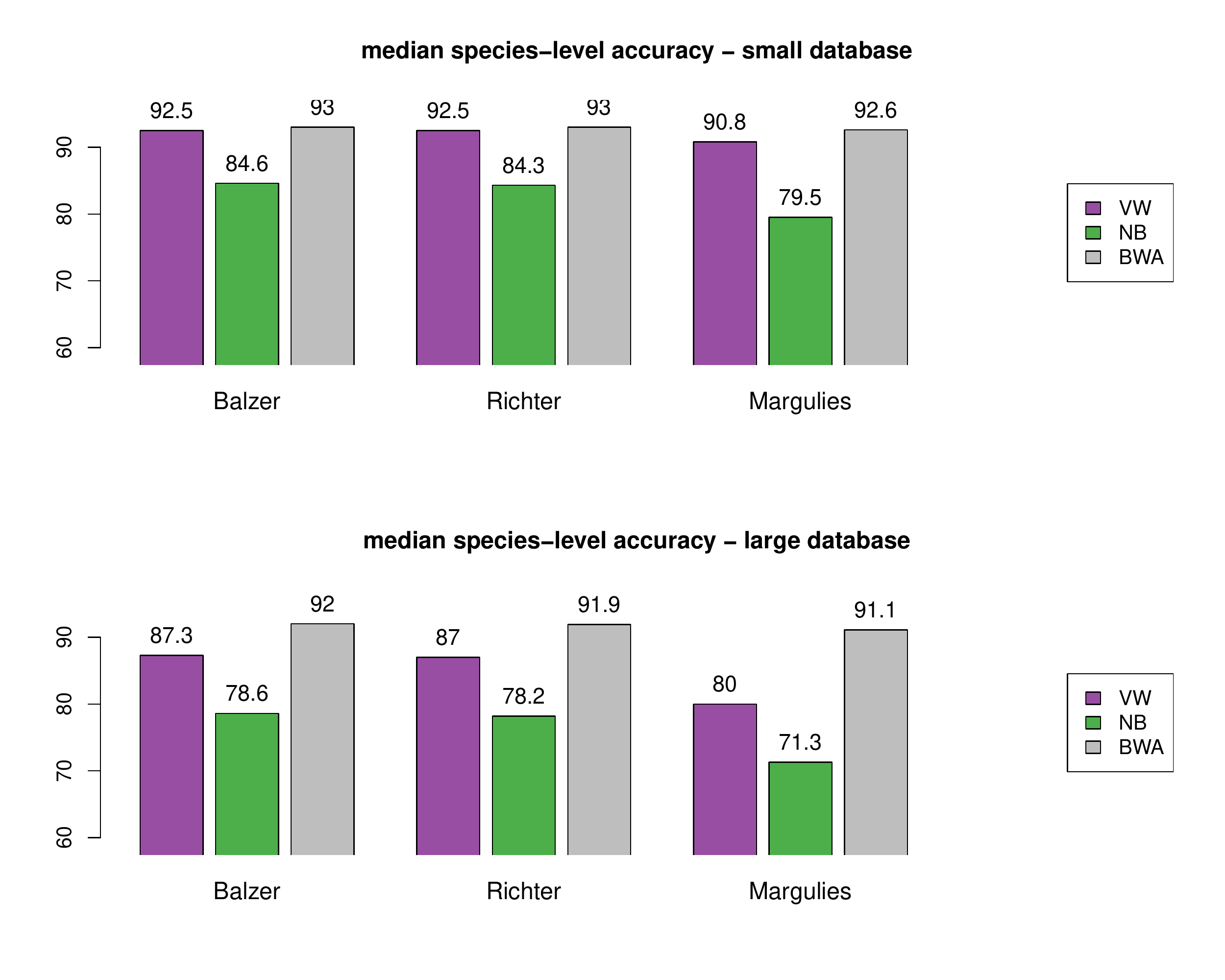}}
	{\caption[\textbf{Robustness to sequencing errors: homopolymer-based models.}]{\textbf{Robustness to sequencing errors: homopolymer-based models.} Top: small reference database. Bottom: large reference database. These figures give the median species-level accuracy obtained with VW (purple), NB (green) and BWA (grey). Each approach has been evaluated on three datasets simulated according to three different error models: Balzer~\citep{balzer2010characteristics}, Richter~\citep{richter2008metasim}, and Margulies~\citep{margulies2005genome}.
		}\label{fig:FCP-homo}}
\end{figure}

To evaluate the impact of homopolymeric errors, we consider the three error models implemented in Grinder : \texttt{Balzer} \citep{balzer2010characteristics}, \texttt{Richter} \citep{richter2008metasim} and \texttt{Margulies} \citep{margulies2005genome}.
Results are shown in Figure \ref{fig:FCP-homo}.
We first note that this kind of errors has a very limited impact on BWA-MEM: only the \texttt{Margulies} model turns out to be detrimental, with a drop of less than 1\% for both the small and large reference databases.
The \texttt{Balzer} and \texttt{Richter} models have a limited impact on the compositional approach as well: a drop of less than $1\%$ is observed as well in most cases (except with the NB classifier, where a drop of almost $2\%$ is observed using the large reference database and the \texttt{Richter} model).
The \texttt{Margulies} model, on the other hand, has a much more severe impact on the performance of $k$-mer based approaches. 
While a relatively limited performance drop of around $1.5\%$ is observed with VW using the small reference database, the NB shows a drop of more than $5\%$.
Considering the large reference database, both approaches show a drop of almost $8\%$, which therefore leads to a gap of more than $10\%$ and up to $20\%$, for VW and NB respectively, compared to the performance of the alignment-based approach. This discrepancy is therefore significantly higher than the one observed from fragments, where VW and NB have performance lower than that of BWA-MEM by around $4\%$ and $12\%$, respectively, on the large reference database.
Analyzing the error profile of the reads obtained by Grinder reveals that both the \texttt{Balzer} and \texttt{Richter} models lead to a median mutation rate of $0.5\%$ (meaning that half of the 200 bp simulated reads show more than one modified base), while this rate raises to $3\%$ with the \texttt{Margulies} model.
While this can readily explain why this latter model had a stronger impact, it suggests that what may be seen as a relatively moderate modification of the sequences (6 bases out of 200) can have a severe impact on compositional approaches.

\subsubsection*{Mutation error model.}
\begin{figure}[!tpb]
	\centerline{\includegraphics[page=1,width=0.6\textwidth]{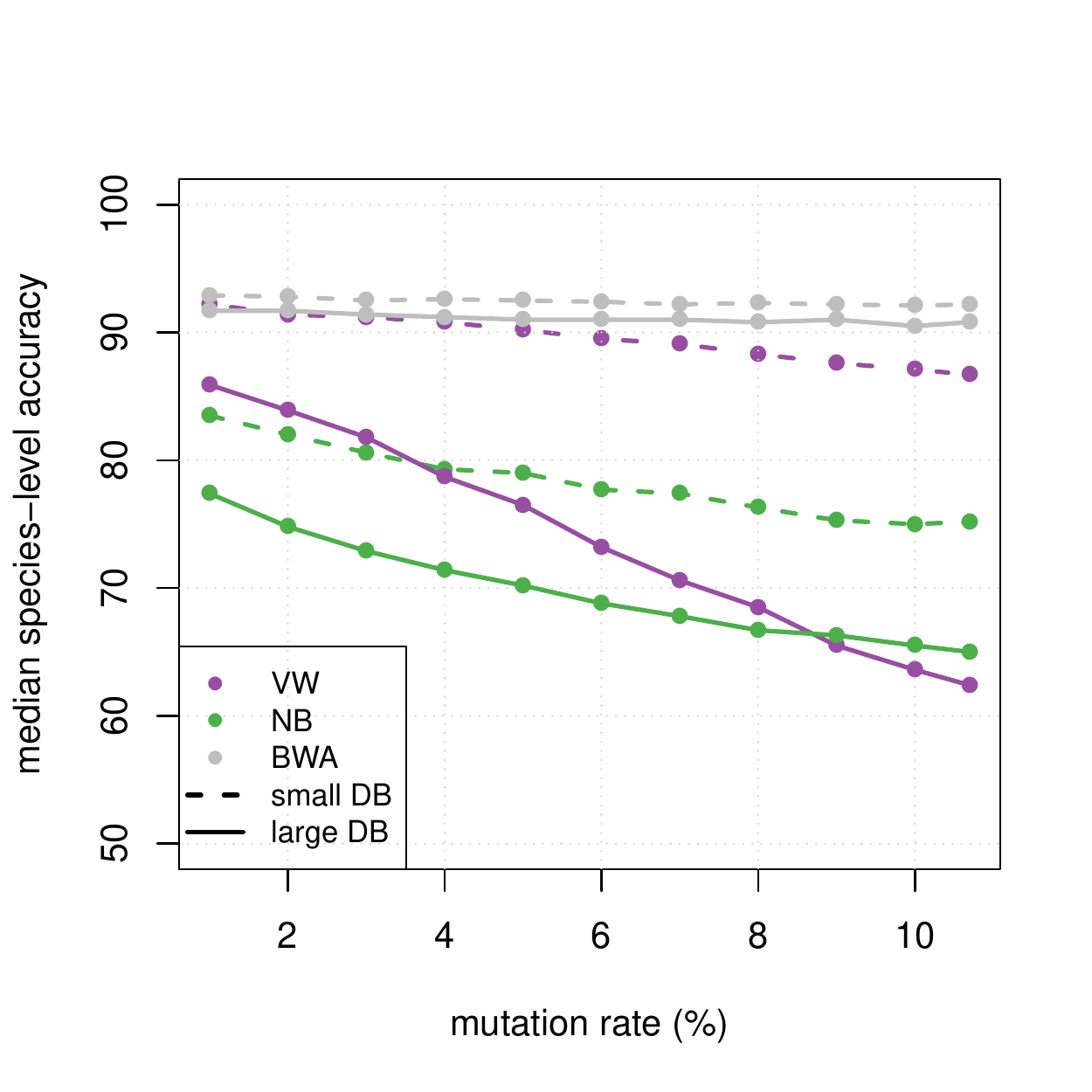}}
	{\caption[\textbf{Robustness to sequencing errors: mutation-based models.}]{\textbf{Robustness to sequencing errors: mutation-based models.} 
			These figures give the median species-level accuracy obtained with VW (purple), NB (green) and BWA (grey), obtained with the small (dashed lines) and large (solid lines) reference databases. Each approach has been evaluated on 11 datasets simulated according to different mutation rates (from 1 to $10.7\%$) with the error model proposed in \citet{korbel2009pemer}.
		}\label{fig:FCP-mutation}}
\end{figure}

To study the impact of general mutation errors, we consider the 4th degree polynomial proposed by \citet{korbel2009pemer} and implemented in Grinder.
Using the default values proposed by Grinder, we empirically observe a median mutation rate of $10.7\%$.
This value is much more important than what is expected by current NGS technologies, and is probably due to the fact that this model was calibrated from shorter reads. Indeed, the median mutation rate decreases to around $1.5\%$ when we reduce the length of the reads to 30, in agreement with the results of the original publication \citep{korbel2009pemer}.
To investigate in details the impact of mutations within reads of length 200, we modify the parameters of the error model in order to gradually increase the median mutation rate from $1\%$ to $10\%$, by $1\%$.
This therefore leads to simulating 11 datasets, since we consider in addition the default Grinder configuration.
Results are shown in Figure \ref{fig:FCP-mutation}.
We first note that this type of errors has a very limited impact on alignment-based approach: even at the higher rate of mutation considered (median mutation rate of $10.7\%$), the performance drops by around $1\%$ with respect to the performance obtained with fragments, for both the small and large reference databases.
On the other hand, the performance obtained with compositional approaches steadily decreases when the mutation rate increases.
Using the small reference database, the impact is more severe for NB than for VW : a drop of up to $10\%$ is observed in the former case (from $85.1\%$ for fragments down to $75.2\%$ for a mutation rate of $10.7\%$) and almost $6\%$ in the latter (from $92.4\%$ down to $86.7\%$).
The drop is even more severe using the larger database in both cases.
Interestingly while it remains relatively constant around $10\%$ for mutation rate greater than $4\%$ with NB (hence twice the gap observed between the small and large datasets using fragments), it keeps increasing with VW and reaches $24\%$ at the highest mutation rate considered. As a result, VW is outperformed by NB using the large reference database for mutation rates greater than $8\%$.
Although these extreme configurations are not realistic regarding the current state of the NGS technologies, we emphasize, in agreement with the previous experiment on homopolymeric errors, that significant drops are observed with compositional approaches for moderate mutation rates, especially for large number of candidate species. For instance, with a mutation rate of $2\%$, the performances of VW and NB drop respectively by $4$ and $5\%$ with the large reference database, while this has no impact on the alignment approach. 
In this more realistic setting, the alignment-based approach shows markedly higher performances: it provides a median species-level accuracy of $91.7\%$, while VW and the NB classifier reach $83.9\%$ $74.8\%$, respectively.

\subsection{Classification speed}
Last but not least, we now turn to the comparison of the comparative and compositional approaches in terms of prediction time. 
This aspect is indeed of critical importance for the analysis of the large volumes of sequence data provided by next-generation sequencing technologies, and constitutes the main motivation of resorting to 
$k$-mer based approaches.
To perform this evaluation we measure the time taken by BWA-MEM and the $k$-mer based approaches to process the 30 test datasets involved in the previous experiments (1 fragments dataset, 3 reads datasets with homopolymeric errors and 11 reads datasets with mutation errors, for the two reference databases considered).
This allows us to investigate the impact of the number of species involved in the reference database, as well as the amount of sequencing noise in the reads.
We do not make a distinction between the two compositional approaches: both involve computing a score for each candidate species, defined as a dot product between the $k$-mer profile of the sequence to classify and a vector of weights obtained by training the model.
To compute this dot-product efficiently, we implemented a procedure described in \citet{sonnenburg2006large}. With this procedure, each A, T, G, C nucleotide is encoded by two bits, which allows to directly convert a $k$-mer as in integer between 0 and $4^k-1$. 
Provided that the weight vector is loaded into memory, the score can be computed ``on the fly" while evaluating the $k$-mer profile of the sequence to be classified, by adding the contribution of the current $k$-mer to the score. 
The drawback of this procedure lies in the fact that the vectors of weights defining the classification models need to be loaded into memory, which can be cumbersome in a multiclass setting.
For 193 and 774 species and $k$-mers of size 12, this amounted to $12$ and $48$ gigabytes, respectively.

\begin{figure}[!tpb]
	\centerline{\includegraphics[width=0.6\textwidth]{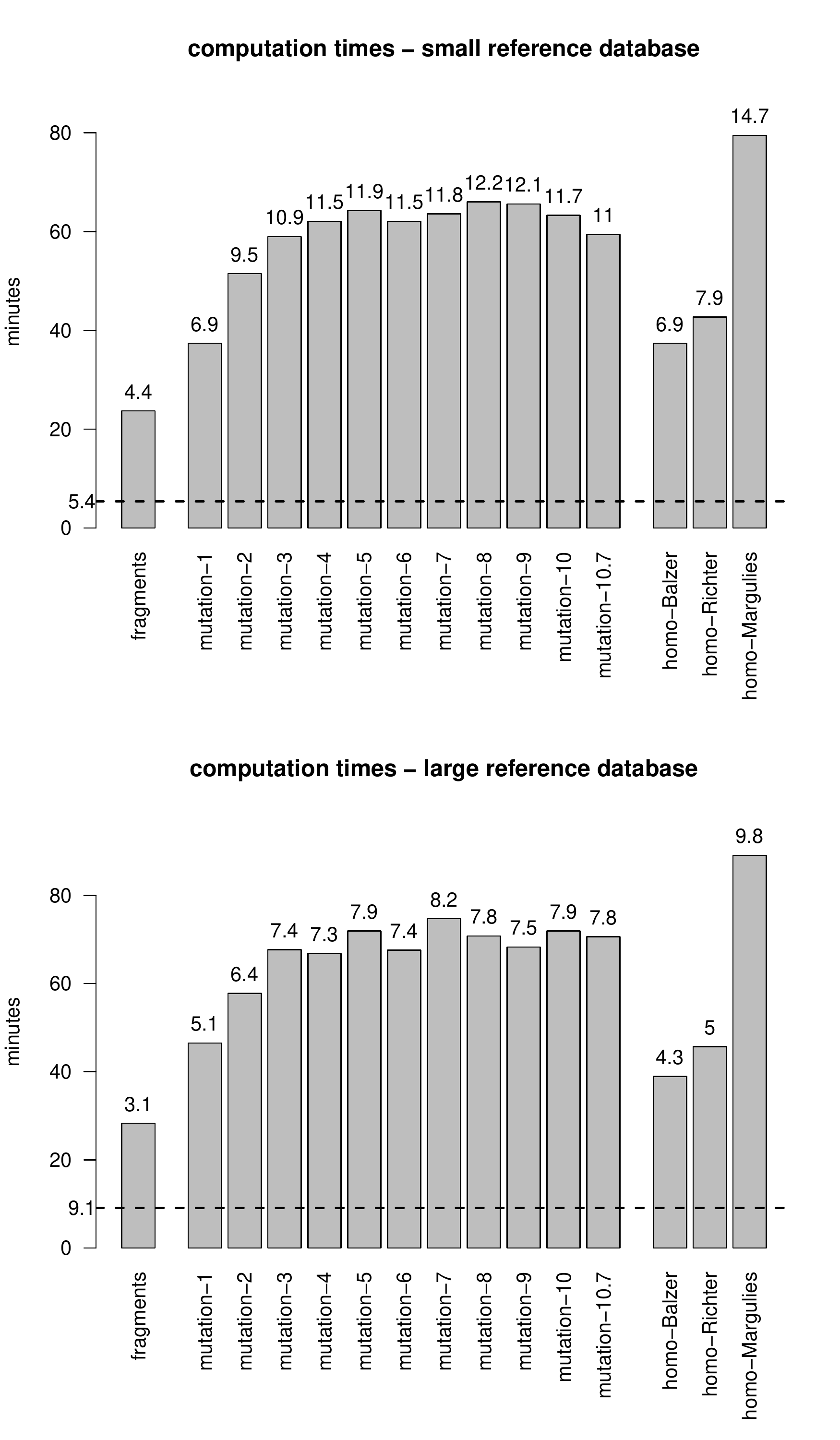}}
	{\caption[\textbf{Evaluation on FCP dataset: mutation-based models.}]{\textbf{Classification times.} The bars represent the time (in minutes) required to classify each fragment or read dataset using BWA-MEM, using the small (top) or large (bottom) reference database. The dashed horizontal lines represent the median time required by VW. The figures shown on top of the bars represent the ratio between the times taken by BWA-MEM and VW.}\label{fig:pred-time}}
\end{figure}

Computation times are measured on a single CPU (Intel XEON - 2.8 Ghz) equipped with 250 GB of memory, and summarized in Figure~\ref{fig:pred-time}.
The time needed to classify each read or fragment dataset by the $k$-mer approach shows little variation, for a given reference database. The median value obtained across test datasets reaches 5.4 and 9.1 minutes, using the small and large reference database, respectively, hence about a two-fold difference. This therefore amounts to classifying around $6.5\times10^5$ and $3.9\times10^5$ reads per minute, respectively.
BWA-MEM shows a different behavior. We observe that the time varies more across reads and fragment datasets, and tends to increase with the amount of sequencing noise. On the other hand, the size of the reference database has a lesser impact, with at most an increase of $20\%$ between the time needed to process a test dataset with the small or large reference databases.
The compositional approach systematically offers shorter prediction times, with an improvement of 3 to almost 15 times, depending on the configuration.

\section{Discussion}\label{sec:discussion}
In this work, we investigate the potential of modern, large-scale machine learning approaches for taxonomic binning of metagenomics data.
We extensively evaluate their performance when the scale of the problem increases regarding (i) the length of the $k$-mers considered to represent a sequence, (ii) the number of fragments used to learn the model, and (iii) the number of candidate species involved in the reference database. 
We also investigate in details their robustness to sequencing errors using simulated reads.
We consider two baselines for this evaluation: a comparative approach based on the BWA-MEM sequence aligner and a compositional approach based on the generative NB classifier. 
We demonstrate in particular that increasing the number of fragments used to train the model has a significant impact on the accuracy of the model, and allows to estimate models based on longer $k$-mers.
While this could be expected and was already highlighted by previous studies, the resulting configurations are out of reach of standard SVM implementations.
We also show that discriminatively trained compositional models usually offer significantly higher performances than generative NB classifiers.
The resulting models are competitive with well-established alignment tools for problems involving a small to moderate number of candidate species, and for reasonable amounts of sequencing errors.
Our results suggest, however, that compositional approaches, both discriminative and generative, are still limited in their ability to deal with problems involving more than a few hundreds species.
In this case, indeed, compositional approaches exhibit lower performance than alignment-based approaches and are much more negatively impacted by sequencing errors.
Finally, we confirm that compositional approach achieve faster prediction times.
This is indeed systematically the case in the various configurations listed above, with predictions obtained 3 to 15 times faster by compositional approaches, and, interestingly, depends on the number of candidate species and the level of sequencing noise.
We emphasize, however, that fast predictions can only be obtained provided that the classification models are loaded in memory, hence for a memory footprint that scales linearly with the number of candidate species and exponentially with the size of the $k$-mers, which can become important for large reference databases and long $k$-mers.

At least three simple extensions could be envisioned to make compositional approaches more competitive in accuracy with the alignment-based approach, faster, and to limit their memory footprint.
First, the robustness to sequencing errors may be improved by learning models from simulated reads instead of fragments. 
This could indeed allow to tune the model to the sequencing technology producing the reads to be analyzed, provided its error model is properly known and characterized.
Second, introducing a sparsity-inducing penalty while learning the model would have the effect of reducing the number of features entering the model, hence to reduce the memory footprint required to load the model into memory.
Finally, alternative strategies, known as error correcting tournaments \citep{beygelzimer2009error}, could be straightforwardly considered to reduce the number of models to learn, hence to store into memory during prediction, to address a multiclass problem. 
Our results indeed suggest that addressing these issues is critical to build state-of-the-art compositional classifiers to analyze metagenomics samples that may involve a broad spectrum of species.
We emphasize however that such large scale models can remain competitive for realistic amounts of sequencing errors and a moderate number of species (around 200 in our study), hence can already be useful in cases where the number of species that can be encountered is limited, which may in particular be the case for diagnostic applications involving specific types of specimens.


\end{document}